\documentclass[prb,reprint,twocolumn,amsmath,amssymb,amsfonts,superscriptaddress,showkeys]{revtex4}

\usepackage{graphicx}
\usepackage{dcolumn}
\usepackage{bm}
\usepackage{color}
\usepackage{amsmath,amsthm}
\usepackage{amssymb,bm}
\usepackage{mathrsfs}
\usepackage{comment}
\usepackage{ulem}

\graphicspath{%
    {converted_graphics/}
    {/}
}
\begin{document}

\preprint{APS/123-QED}


\title{Control of Excitation Energy Transfer in Condensed Phase Molecular Systems by Floquet Engineering}
\author{Nguyen Thanh Phuc}
\affiliation{Department of Theoretical and Computational Molecular Science, Institute for Molecular Science, Okazaki 444-8585, Japan}
\affiliation{Department of Structural Molecular Science, The Graduate University for Advanced Studies, Okazaki 444-8585, Japan}
\email{nthanhphuc@ims.ac.jp}
\author{Akihito Ishizaki}
\affiliation{Department of Theoretical and Computational Molecular Science, Institute for Molecular Science, Okazaki 444-8585, Japan}
\affiliation{Department of Structural Molecular Science, The Graduate University for Advanced Studies, Okazaki 444-8585, Japan}

\begin{abstract}
\section{Abstract}
Excitation energy transfer (EET) is one of the most important processes in both natural and artificial chemical systems including, for example, photosynthetic complexes and organic solar cells. The EET rate, however, is strongly suppressed when there is a large difference in the excitation energy between the donor and acceptor molecules. Here, we demonstrate both analytically and numerically that the EET rate can be greatly enhanced by periodically modulating the excitation energy difference. The enhancement of EET by using this Floquet engineering, in which the system's Hamiltonian is made periodically time-dependent, turns out to be efficient even in the presence of strong fluctuations and dissipations induced by the coupling with a huge number of dynamic degrees of freedom in the surrounding molecular environments. As an effect of the environment on the Floquet engineering of EET, the optimal driving frequency is found to depend on the relative magnitudes of the system and environment's characteristic time scales with an observed frequency shift when moving from the limit of slow environmental fluctuations (inhomogeneous broadening limit) to that of fast fluctuations (homogeneous broadening limit). 
\end{abstract}


\maketitle

\section{Graphical TOC}

\begin{figure}[h] 
  \centering
  \includegraphics[width=5cm,keepaspectratio]{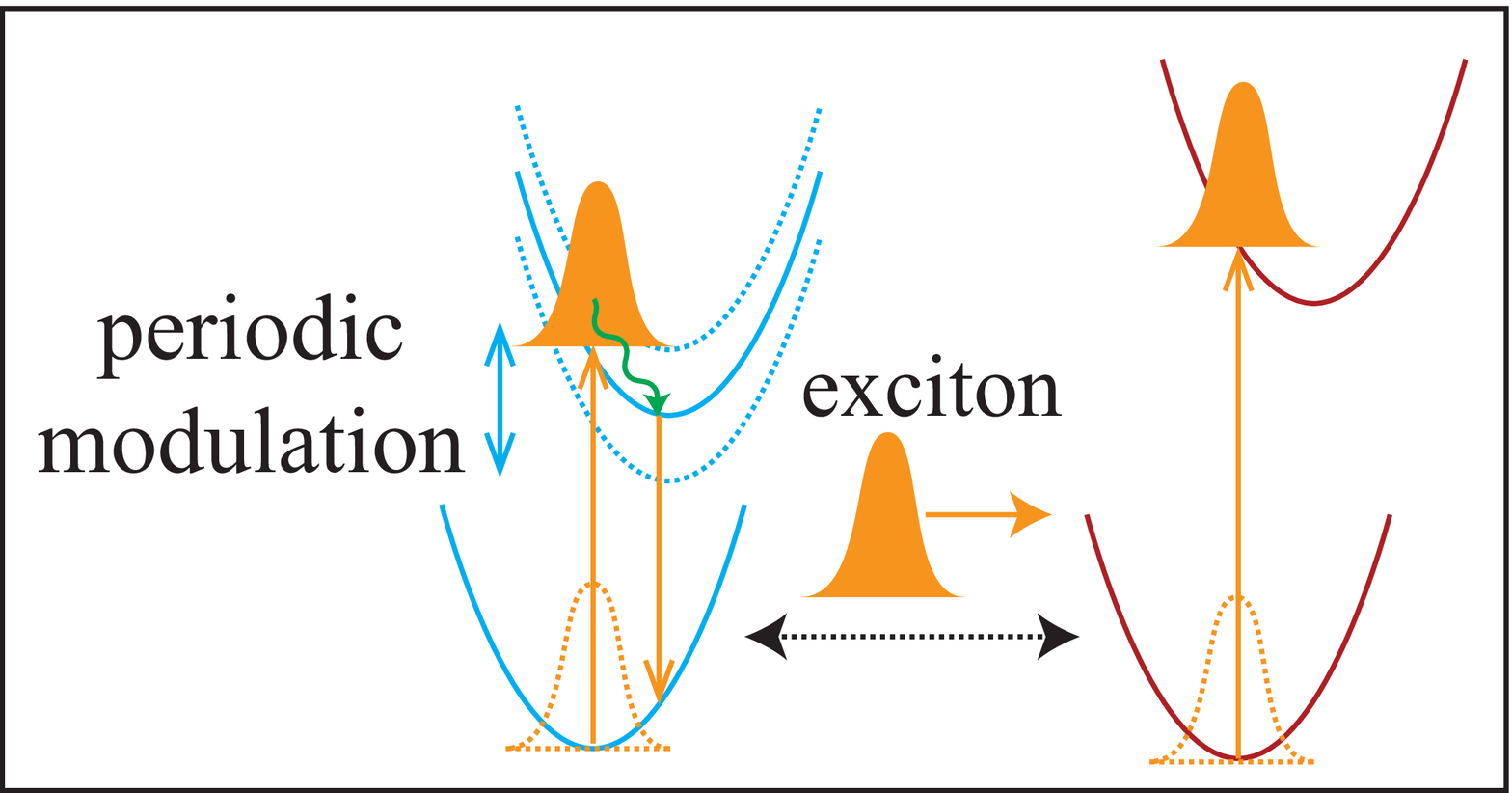}
\end{figure}

Excitation energy transfer (EET) is one of the most elementary and vital chemical processes in molecular systems.
For example, EET from the light-harvesting antennae to the reaction centers in photosynthetic organisms is crucial for understanding their extremely high quantum efficiency under low light conditions.
\cite{Blankenship-book, Scholes11, Mirkovic17}
Under high light conditions, on the other hand, photosynthetic systems regulate the EET so that the amount of electronic excitations does not exceed the capacity of the reaction centers
\cite{Horton96, Croce14}
Excitation energy transfer is also an indispensable process in the working of photovoltaic systems such as organic solar cells, where Frenkel excitons are transported to the bulk heterojunction's interface between the electron-donor domain of conjugated polymers and the electron-acceptor domain of fullerenes at which charge transfer occurs to produce electron--hole pairs. \cite{Blom07, Collini09, Savoie14} It is expected that understanding of the EET in natural photosynthetic systems can be exploited to improve the energy conversion efficiency in the photovoltaics. \cite{Bredas17, Green17}
A great deal of effort has been made to understand EET processes in complex molecular systems; however, recent advances in optical and spectroscopic technologies have added new dimensions to the investigation. 

Recently, two-dimensional electronic spectroscopy revealed the existence of long-lived quantum coherence among the electronic excitations of pigments embedded in light-harvesting proteins of different types of biological organisms.~\cite{Engel07, Lee07, Calhoun09, Collini10,Panitchayangkoon10, Schlau-Cohen12, Westenhoff12,Romero14, Fuller14,Scholes17} The interplay of the quantum coherence of the electronic excitation in pigments and the thermal fluctuations arising from the surrounding protein environment in the EET of photosynthesis has been extensively investigated. \cite{Mohseni08, Plenio08, Jang08, Ishizaki09, Cao2009, Wu12, Huelga13, Fassioli14, Chenu15, Ishizaki12} 
However, the most fundamental factors that determine the pathways of EET and the EET efficiency are still the strengths of electrostatic interactions among molecules and the energy landscape of the involved electronic excitations.\cite{Renger09, Yang02} Indeed, when the excitation energy difference between the donor and acceptor molecules is large compared with the other relevant energy scales of the system under consideration, the EET rate is strongly suppressed, as F\"orster theory demonstrates.\cite{Forster46, Yang02}
In this case, the so-called Floquet engineering \cite{Grifoni98, Eckardt17, Holthaus16, Bukov15} can be exploited to enhance the EET rate. In Floquet engineering, the system Hamiltonian is periodically modulated with a frequency $\omega$. When the excitation energy difference $\Delta E$ between the donor and the acceptor is close to an integer multiple of the driving frequency, $\Delta E\simeq n\hbar\omega$ with $n$ being an integer, the EET rate would be greatly enhanced via absorption (emission) of $n$ energy quanta from the driving source for negative (positive) $\Delta E$. Floquet engineering has been investigated for controlling 
transport of ultracold atoms in shaken optical lattices. \cite{Sias08, Ivanov08} In contrast to ultracold atoms, which are almost isolated quantum systems, however, electronic excitations in condensed phase molecular systems are often strongly coupled to their surrounding environments, and hence, they are substantially affected by random fluctuations as well as molecular vibrational motions. Similar ideas have been applied in different types of systems including electron transfer in molecules \cite{Dakhnovskii95} and electron transport in nanostructured devices. \cite{Kohler05}

In this Letter, we investigate how the EET in condensed phase molecular systems can be controlled through the use of Floquet engineering. Here, the difference in the excitation energy between the donor and acceptor is periodically modulated by taking advantage of the difference between the two molecules. If a molecule possesses a permanent dipole moment (PDM), the periodic modulation of its excitation energy can be generated by applying an electromagnetic field (EMF) that interacts with the molecule via the dipole interaction. In contrast, if the molecule does not have a PDM, the periodic modulation of the excitation energy can be realized, for example, through the ac Stark effect with the amplitude of the applied EMF being varied periodically or by perturbing the surrounding environment of the molecule with an oscillation through the piezoelectric effect. We demonstrate, both analytically by using the general Floquet theory with a high-frequency approximation and numerically by solving the hierarchy equation of motion, that the EET rate can be greatly enhanced with the use of Floquet engineering and that this enhancement is efficient even in the presence of strong dissipations and fluctuations induced by coupling with a huge number of dynamic degrees of freedom of the surrounding environment. On the other hand, the effect of molecular environment on Floquet engineering of EET is also unveiled. The optimal driving frequency is found to depend on the relative magnitudes of the system and the environment's characteristic time scales. In particular, we find an observable shift in the frequency when moving from the limit of slow nuclear motion, i.e., the inhomogeneous broadening limit, to that of fast nuclear motion, i.e., the homogeneous broadening limit. 

\begin{figure}[tbp] 
  \centering
  \includegraphics[width=3in,keepaspectratio]{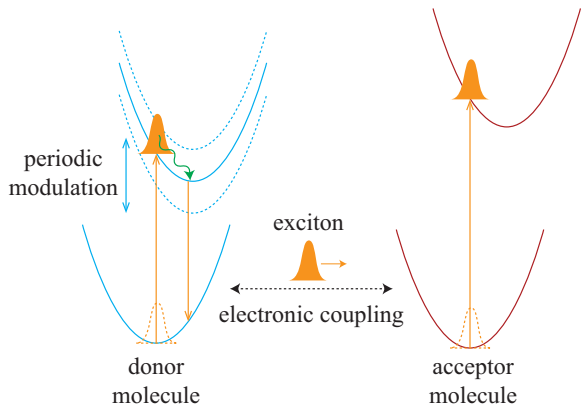}
  \caption{EET between two molecules in condensed phases: the donor (blue) and the acceptor (red). Each molecule has ground-state and excited-state PESs that are modeled by two harmonic potentials. Their equilibrium positions are shifted from each other as a consequence of the coupling between the electronic and nuclear degrees of freedom in the molecule. An electronic excitation, i.e., exciton (yellow), is generated in the donor molecule through a vertical Franck--Condon transition. After dissipating energy to the nuclear environment and approaching the equilibrium position in the excited-state PES, the exciton is transferred to the acceptor molecule via the electronic coupling between the two molecules. The EET rate can be greatly enhanced by periodically modulating the difference in the excitation energy between the donor and acceptor.}
  \label{fig: energy levels}
\end{figure}

We consider EET between two molecules in condensed phases, each of which are coupled with the environmental degrees of freedom. The potential energy surfaces (PESs) for the ground and excited states of the donor and acceptor molecules are approximated by harmonic potentials. Initially, the donor molecule is prepared in the excitation state, for example, by absorbing a short light pulse. The excitation is assumed to occur vertically, i.e., without changing the nuclear configuration, following the Franck--Condon approximation, as shown in Figure~\ref{fig: energy levels}. The excitation energy transfer is performed by an electronic coupling $V_{12}$ between the donor and acceptor molecules. The Hamiltonian of the total system is given by~\cite{Ishizaki12} 
\begin{align}
	H=&
	\sum_{m=1}^2\sum_{a=\mathrm{g},\mathrm{e}} 
	H_{ma}(\mathbf{x}_m) \lvert\varphi_{ma}\rangle\langle\varphi_{ma}\rvert
	\nonumber\\
	&+(\hbar V_{12}\lvert\varphi_{1\mathrm{e}}\rangle\langle\varphi_{1\mathrm{g}}\rvert
	\otimes
	\lvert\varphi_{2\mathrm{g}}\rangle\langle\varphi_{2\mathrm{e}}\rvert+ \text{h.c.}), 
	\label{eq: static Hamiltonian}
\end{align}
where $|\varphi_{ma}\rangle$ and $H_{ma}(\mathbf{x}_m)$ ($m=1, 2$, $a=\mathrm{g},\mathrm{e}$) represent the electronic state vectors and the Hamiltonians describing the nuclear dynamics associated with the electronic ground (g) and excited (e) states of the molecules, respectively. Here, $\mathbf{x}_m$ represents the set of relevant nuclear coordinates including the normal modes of the intramolecular vibrations and the surrounding environment. The typically small dependence of the electronic coupling $V_{12}$ on the nuclear degrees of freedom is neglected. The Hamiltonian $H_{ma}(\mathbf{x}_m)$ is a sum of the nuclear kinetic energy and the PES $\epsilon_{ma}(\mathbf{x}_m)$:
\begin{align}
	H_{m\mathrm{g}}(\mathbf{x}_m)
	&=\epsilon_{m\mathrm{g}}(\mathbf{x}_{m\mathrm{g}}^0)+\sum_\xi \frac{\hbar\omega_{m\xi}}{2}(p_{m\xi}^2+q_{m\xi}^2),
	\\
	H_{m\mathrm{e}}(\mathbf{x}_m)
	&=H_{m\mathrm{g}}(\mathbf{x}_m)+\hbar\Omega_m-\sum_\xi \hbar\omega_{m\xi} d_{m\xi}q_{m\xi},
\label{eq: excited state PES}
\end{align}
where $\mathbf{x}_{m\mathrm{g}}^0$ is the equilibrium configuration of the nuclear coordinates associated with the electronic ground state, $q_{m\xi}$ and $p_{m\xi}$ are the dimensionless coordinate and momentum, respectively, of the normal mode $\xi$ with the corresponding frequency $\omega_{m\xi}$, and $d_{m\xi}$ is the dimensionless displacement, i.e., the distance between the respective equilibrium configurations $\mathbf{x}_{m\mathrm{g}}^0$ and $\mathbf{x}_{m\mathrm{e}}^0$ of the nuclear coordinate associated with the electronic ground and excited states. In the following, we set the origin of energy $\epsilon_{m\mathrm{g}}(\mathbf{x}_{m\mathrm{g}}^0)=0$ without loss of generality. The Franck--Condon transition energy is given by $\hbar\Omega_m \equiv \epsilon_{m\mathrm{e}}(\mathbf{x}_{m\mathrm{g}}^0)-\epsilon_{m\mathrm{g}}(\mathbf{x}_{m\mathrm{g}}^0)$. The reorganization energy $\hbar\lambda_m\equiv\epsilon_{m\mathrm{e}}(\mathbf{x}_{m\mathrm{g}}^0)-\epsilon_{m\mathrm{e}}(\mathbf{x}_{m\mathrm{e}}^0)$ is the dissipating energy when the nuclear configuration changes from $\mathbf{x}_{m\mathrm{g}}^0$ after the vertical Franck--Condon transition to its equilibrium point $\mathbf{x}_{m\mathrm{e}}^0$ in the excited-state PES.

When the intermolecular electronic coupling is weak compared to the inverse time scale of the molecule's reorganization, $V_{12}\ll\gamma_m\equiv\tau_m^{-1}$, i.e., in the incoherent EET regime, the environmental degrees of freedom associated with the donor would relax to its equilibrium configuration in the excited-state manifold before the excitation is transferred to the acceptor. Using Fermi's golden rule, we obtain the F\"orster formula for the EET rate from state $\lvert 1 \rangle = \lvert \varphi_{1\mathrm{e}}\rangle\lvert\varphi_{2\mathrm{g}}\rangle$ to state $\lvert 2 \rangle = \lvert \varphi_{1\mathrm{g}}\rangle\lvert\varphi_{2\mathrm{e}}\rangle$,
$k_{1\to 2}= (|V_{12}|^2/2\pi) \int_{-\infty}^\infty {\rm d}\omega\, F_1(\omega)A_2(\omega)$,
where $F_1(\omega)$ and $A_2(\omega)$ are the fluorescence line shape of the donor and absorption line shape of the acceptor, respectively.  \cite{Forster46, Yang02} In the inhomogeneous broadening limit, the fluorescence and absorption line shapes may be approximated by the classical Gaussian forms, and consequently,
the EET rate is given by \cite{May-book, Cleary13}
\begin{align}
	k_{1\to 2}^\mathrm{inh}
	=
	|V_{12}|^2
	\sqrt{\frac{\pi\hbar}{k_\mathrm{B}T(\lambda_1+\lambda_2)}}
	\exp\left[
		-\frac{\hbar(\Omega_1-2\lambda_1-\Omega_2)^2}{4k_\mathrm{B}T(\lambda_1+\lambda_2)}
	\right]
\end{align}
In the opposite limit, i.e., the homogeneous broadening limit, the line shapes have Lorentzian forms, and hence, the EET rate is given by \cite{May-book, Cleary13}
\begin{align}
	k_{1\to 2}^\mathrm{hom}
	=
	\frac{2  |V_{12}|^2 \Gamma_{12}}{(\Omega_1-\Omega_2)^2+\Gamma_{12}^2},
\end{align}
where $\Gamma_{12}\equiv (2k_\mathrm{B}T/\hbar)[(\lambda_1/\gamma_1)+(\lambda_2/\gamma_2)]$. In both the inhomogeneous and homogeneous limits, the EET rate would be strongly suppressed when there is a large difference in the excitation energy between the donor and acceptor molecules.

Now, let us consider a case that the difference in the excitation energy between the donor and acceptor is periodically modulated with frequency $\omega$ and amplitude $A$. The Hamiltonian of the total system is then given by eq~\eqref{eq: static Hamiltonian} with the excitation energy $\Omega_1$ in eq~\eqref{eq: excited state PES} replaced by a time-dependent one 
\begin{align}
	\Omega_1  \rightarrow \Omega_1(t)=\Omega_1^0+A\cos\omega t.
\end{align}
The driving frequency $\omega$ is chosen such that the static excitation energy difference between the two molecules is close to an integer multiple of $\omega$, $\Omega_2-\Omega_1^0\simeq n \omega$, where $n$ is an integer. Following the general Floquet theory,
\cite{Eckardt17, Engelhardt13}
we perform a unitary transformation defined by the operator
\begin{align}
	U(t)
	=
	e^{
	-i\{ [A\sin(\omega t)/\omega] \lvert\varphi_{1\mathrm{e}}\rangle\langle\varphi_{1\mathrm{e}}\rvert
	+
	n\omega t \lvert\varphi_{2\mathrm{e}}\rangle\langle\varphi_{2\mathrm{e}}\rvert \}
	}.
\end{align}	
The transformed Hamiltonian
\begin{align}
	H'(t)=U^\dagger(t)H(t)U(t)+i\hbar \frac{{\rm d}U^\dagger(t)}{{\rm d}t} U(t),
\end{align}
has the same form as the original static one, eq~\eqref{eq: static Hamiltonian}. However, the excitation energies of the donor and acceptor are changed to $\Omega_1^0$ and $\Omega_2-n\omega$, respectively.
More importantly, the electronic coupling between the two molecules becomes time-dependent $V_{12}e^{i\chi(t)}$, where $\chi(t)=A\sin(\omega t)/\omega-n\omega t$ (see the Supporting Information for details). Therefore, the whole time dependence of the transformed Hamiltonian is contained in the phase factor $e^{i\chi(t)}$ of the electronic coupling.

In the high-frequency limit, i.e., when the driving frequency is large compared with the characteristic energy scales relevant to the EET, we can take an average over the rapid oscillation in the Hamiltonian.
\cite{Eckardt17}
As a result, we obtain an effective Hamiltonian approximated by its cycle average
\begin{align}
	H^\mathrm{eff} = \frac{1}{T}\int_0^T \text{d}t\, H'(t),
\end{align} 
where $T=2\pi/\omega$ is the period of the driving. This averaged Hamiltonian corresponds to the lowest-order 
term in the high-frequency expansion, i.e., an expansion in powers of $1/\omega$. The time-independent effective Hamiltonian then has the same form as the static one (eq~\eqref{eq: static Hamiltonian}), with the excitation energies of the donor and acceptor being $\Omega_1^0$ and $\Omega_2-n\omega$, respectively.
Furthermore, the time-independent effective electronic coupling is given by 
\begin{align}
	V_{12}^\mathrm{eff}
	=
	\frac{V_{12}}{T}\int_0^T \text{d}t\, e^{i\chi(t)}=V_{12}J_n(A/\omega),
\end{align} 
where $J_n(x)$ is the $n$th-order Bessel function of the first kind, $J_n(x)=(1/2\pi) \int_0^{2\pi}\text{d}t\,e^{i(-n t+x\sin t)}$. Compared with the original Hamiltonian in eq~\eqref{eq: static Hamiltonian}, the effective Hamiltonian in the presence of a periodic driving has a reduced excitation-energy difference between the donor and the acceptor molecules while at the same time the electronic coupling is modified by the Bessel function. The maximum values of the low-order Bessel functions $J_n(x)$ ($n=1,2,\dots$) are of the order of unity. Because the system--environment couplings in the effective Hamiltonian are the same as those in the original Hamiltonian, the expressions for the EET rate in both the inhomogeneous and homogeneous limits can be applied to the EET with Floquet engineering.

Considering the incoherent EET regime, for the inhomogeneous limit, we choose the optimal driving frequency 
so as to satisfy $\Omega_2-\Omega_1^0+2\lambda_1=n\omega$. In this case, the EET rate with the Floquet engineering is found to be 
\begin{align}
	k_{1\to 2}^\mathrm{inh, Floq}
	=
	|V_{12}|^2J_n(A/\omega)^2
	\sqrt{\frac{\pi\hbar}{k_\mathrm{B}T(\lambda_1+\lambda_2)}}.
\end{align}
In the homogeneous limit, on the other hand, the optimal driving frequency should be chosen as $\Omega_2-\Omega_1^0=n\omega$. In this case, the Floquet-engineered EET rate is given by 
\begin{align}
	k_{1\to 2}^\mathrm{hom, Floq}
	=
	\frac{2|V_{12}|^2J_n(A/\omega)^2}{\Gamma_{12}}.
\end{align}
The optimal driving frequency, therefore, is shifted as moving from the inhomogeneous to the homogeneous broadening limit. This demonstrates an effect of the system's environment on the control of the system's dynamics by the Floquet engineering. It is, however, clear that in both the inhomogeneous and homogeneous limits the EET rate can be greatly enhanced as there are no longer large suppression factors associated with the excitation energy difference between the donor and acceptor. 

Until this point, we have addressed the incoherent hopping regime, the high-frequency limit, and the inhomogeneous and homogeneous broadening limits in order to get insight into the inner working of the Floquet engineering by deriving the analytical expressions for the Floquet-engineered EET rate. To demonstrate that Floquet engineering can be efficient for enhancement of EET over a wide range of parameters, we perform quantum dynamics calculations of the EET.
It should be noticed that quantum dynamics described by the above Hamiltonians can be solved in a numerically accurate fashion \cite{Ishizaki09,Kato:2013bd} through the use of the so-called hierarchical equation of motion approach. \cite{Tanimura89,Tanimura06} 
In this approach, the reorganization energy and  the time scale of the environment-induced fluctuations in the excitation energy of the $m$th molecule are characterized by the relaxation function, $\Psi_m(t) = (2/\pi)\int^\infty_0 \mathrm{d}\omega\,[J_m(\omega) /\omega] \cos\omega t$, where $J_m(\omega)$ stands for the spectral density.
Specifically, when the spectral density is given by the Drude--Lorentz form, $J_m(\omega) = 2\hbar\lambda_m \gamma_m \omega/(\omega^2+\gamma_m^2)$, the relaxation function is expressed as $\Psi_m(t) = 2\lambda_m \exp(-\gamma_m t)$.
To focus on roles of the reorganization energy $\hbar\lambda_m$ and the time scale $\gamma_m^{-1}$, therefore, we employ the Drude--Lorenz model in this work.

\begin{figure}[tbp]
\centering
\includegraphics[width=3in, keepaspectratio]{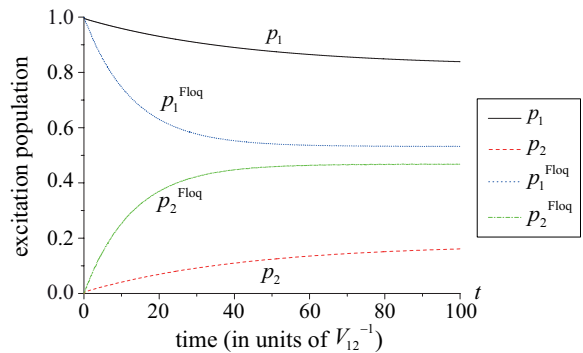}
\caption{Time evolutions of the populations of excitation in the donor and acceptor molecules: without periodic modulation of the excitation energy, $p_1(t)$ (black, solid) for the donor and $p_2(t)$ (red, dashed) for the acceptor; with periodic modulation of the excitation energy, $p_1^\mathrm{Floq}(t)$ (blue, dotted) for the donor and $p_2^\mathrm{Floq}(t)$ (green, dashed dotted) for the acceptor. Here, the time $t$ is measured in units of the electronic coupling $V_{12}^{-1}$. Values of the parameters are shown in the text.}
\label{fig: time evolution of exciton populations}
\end{figure}

To demonstrate effects of Floquet engineering on the EET rate, we take a large excitation energy difference between the donor and the acceptor. On the other hand, the magnitude of the electronic coupling $V_{12}$ is taken to be comparable to the reorganization energy $\lambda_{1,2}$ as well as the inverse of the environmental relaxation time $\gamma_{1,2}$, for which the system is in the intermediate regime between 
the coherent and incoherent limits. Values of the parameters are taken to be $V_{12}=20\,{\rm cm^{-1}}$, $\Omega_2-\Omega_1^0=600\,{\rm cm^{-1}}$, $\gamma_1=\gamma_2=\lambda_1=\lambda_2=40\,{\rm cm^{-1}}$, and $T=100\,{\rm K}$. 
The driving frequency $\omega=18\,{\rm THz}$ is chosen to be near resonant, and the driving amplitude is chosen to satisfy $A=\omega$. 
Figure~\ref{fig: time evolution of exciton populations} clearly demonstrates that Floquet engineering enhances the EET rate.

\begin{figure}[tbp] 
  \centering
  \includegraphics[width=3in,keepaspectratio]{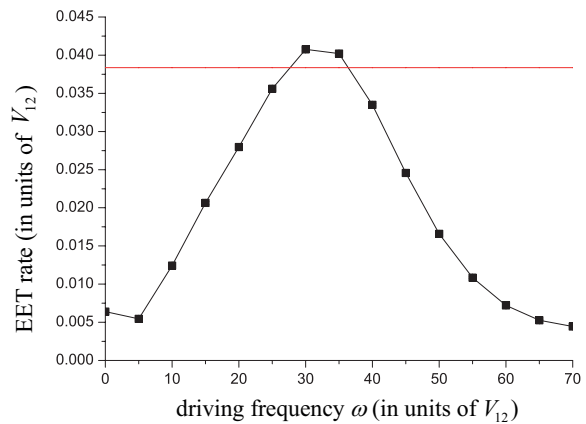}
  \caption{Dependence of the EET rate (black squares) on the driving frequency, both of which are measured in units of the electronic coupling $V_{12}$. The ratio  of the driving amplitude $A$ to the frequency $\omega$ is kept constant, $A/\omega=1$. The red horizontal line indicates the maximum value of EET rate predicted by the analytical expression that is valid in the incoherent EET regime and the inhomogeneous broadening limit.}
  \label{fig: varying driving frequency}
\end{figure}

\begin{figure}[tbp] 
  \centering
  \includegraphics[width=3in,keepaspectratio]{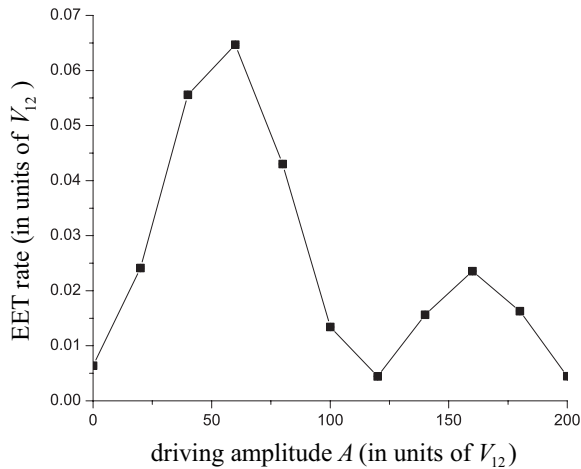}
  \caption{Dependence of the EET rate on the driving amplitude, both of which are measured in units of the electronic coupling $V_{12}$. The driving frequency is fixed to be $\omega=30 V_{12}$.}
  \label{fig: varying driving amplitude}
\end{figure}

To investigate how the EET rate depends on the driving frequency and amplitude, we evaluate the EET rate $k_{1\to 2}$ by numerically calculating the evolution of the reduced density operator $\hat\rho(t)$ with the initial condition $p_1(0)=\langle 1 \vert \hat\rho(0) \vert 1 \rangle = 1$ and $p_2(0)=\langle 2 \vert \hat\rho(0) \vert 2 \rangle = 0$. In general, EET rates are defined by the rate equations ${\rm d}p_1(t)/{\rm d}t=-k_{1\to 2}p_1(t)+k_{2\to 1}p_2(t)$ and ${\rm d}p_2(t)/{\rm d}t=-k_{2\to 1}p_2(t)+k_{1\to 2}p_1(t)$. 
At short time $t\simeq 0$, the equations are approximated by ${\rm d}p_1(t)/{\rm d}t\simeq -k_{1\to 2}p_1(t)$ because of $p_2(t)\simeq0$, and thus, the EET rate can be obtained by fitting the short-time evolution of $p_1(t)$ according to $p_1(t)\simeq p_1(0)\exp({-k_{1\to 2}t})\simeq 1-k_{1\to 2}t$. Figure~\ref{fig: varying driving frequency} presents the EET rate as a function of the driving frequency $\omega$ with keeping the ratio $A/\omega$  constant. It is evident that the EET rate is maximum near the resonance frequency $\Omega_2-\Omega_1^0=30 V_{12}$. The numerically obtained maximum value of the EET rate is found to be slightly larger than the analytically predicted maximum value for the incoherent EET regime and the inhomogeneous broadening limit, $k^\mathrm{max}=0.03837V_{12}$, which is colored in red. 
The dependence of the EET rate on the driving amplitude when the driving frequency is kept constant at 
$\omega=30 V_{12}$ is shown in Figure~\ref{fig: varying driving amplitude}. 
The observed oscillating behavior of $k_{1\to 2}$ as a function of $A$ originates from the pattern of the first-order Bessel function $J_1(x)$.

\begin{figure}[tbp] 
  \centering
  \includegraphics[width=3in,keepaspectratio]{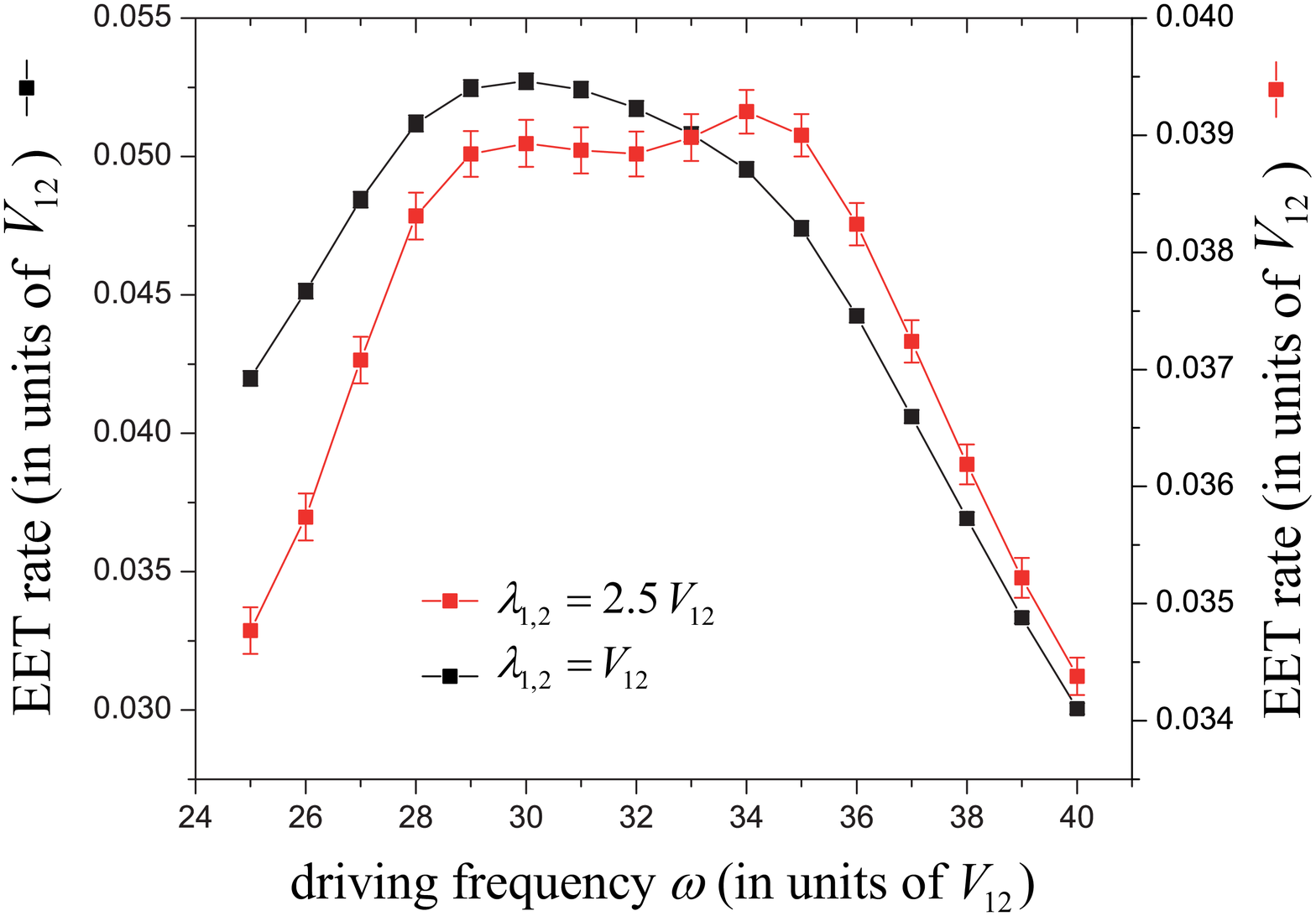}
  \caption{EET rate as a function of the driving frequency for two different values of the reorganization energy: $\lambda_{1,2}=V_{12}$ (black) and $2.5 V_{12}$ (red), where $V_{12}$ is the electronic coupling between the donor and acceptor molecules. The values of the EET rate are shown on the left (right) vertical axis for black (red) data. Here, both the EET rate and the driving frequency are measured in units of $V_{12}$. The error bars are obtained from fitting of the short-time evolution of the excitation population $p_1(t)$ of the donor molecule.}
  \label{fig: slow vs fast nuclear motion}
\end{figure}

In order to reveal the difference in the behavior of EET rate as a function of the driving frequency $
\omega$ in the inhomogeneous and homogeneous broadening limits, we evaluate $k_{1\to 2}(\omega)$ for two different values of the reorganization energy: $\lambda_{1,2}=V_{12}$ and $\lambda_{1,2}=2.5 V_{12}$. The result of $k_{1\to 2}(\omega)$ for the frequency range around the resonance value is shown in Figure~\ref{fig: slow vs fast nuclear motion}. It can be seen that for the larger value of $\lambda_{1,2}$, where the system is closer to the inhomogeneous broadening limit, there appear two peaks of the EET rate around $\Omega_2-\Omega_1^0+2\lambda_1=34 V_{12}$ and $\Omega_2-\Omega_1^0=30 V_{12}$, with the first peak being slightly larger. The emergence of the peak around $\omega=34 V_{12}$, i.e., Stokes-shifted from the resonance frequency $\omega=30 V_{12}$, can be regarded as a consequence of the system's environment in the inhomogeneous limit. This is qualitatively consistent with the analytical prediction in the incoherent EET regime even though the parameters used in the numerical calculation correspond to the intermediate regime between coherent and incoherent hopping. Moreover, it is clear from Figure~\ref{fig: slow vs fast nuclear motion} that as the reorganization energy gets smaller, for which the system moves toward the homogeneous broadening limit, the Stokes-shifted peak is damped while the resonance frequency peak is enhanced. This behavior of $k_{1\to 2}(\omega)$ when $\lambda_{12}$ is varied is also in qualitative consistency with the analytical prediction.

In conclusion, we have demonstrated both analytically and numerically that the EET can be significantly enhanced with the use of Floquet engineering. The enhancement of the EET process by Floquet engineering is found to be efficient even in the presence of fluctuations and dissipations that are induced by coupling with a huge number of dynamic degrees of freedom in the surrounding molecular environments. Floquet engineering, therefore, may provide us with a powerful tool for controlling quantum dynamics in molecular systems in addition to the laser-pulse-shaping approaches that have been widely considered for molecular systems. \cite{Shapiro-book, Rice-book, Rabitz00, Walmsley03}
Unlike quantum control of energy flow based on a shaped laser pulse, \cite{Herek02} which generates a temporally separated sequence of coherent wavepackets that interfere in a manner that enhances the EET, Floquet engineering directly targets the EET of a specific pair of molecules in a molecular system. Furthermore, in addition to the transition amplitude, Floquet engineering can also be used to manipulate the quantum phase coherence of electrons involved in the chemical processes, by which quantum mechanical properties of the molecular system can be explored.

\begin{acknowledgements}
This work was supported by JSPS KAKENHI Grant Number 17H02946 and JSPS KAKENHI Grant Number 17H06437 in Innovative Areas ``Innovations for Light-Energy Conversion (I$^4$LEC)''. N.T.P. thanks Dr.~Akihito Kato for fruitful discussions. 
\end{acknowledgements}


\end{document}